\documentclass[preprint,showpacs,preprintnumbers,amsmath,amssymb]{revtex4}

\usepackage{graphicx}
\usepackage{dcolumn}
\usepackage{bm}

\begin{document}

\preprint{APS/123-QED}

\title{Quantum Hall effect in multilayered massless Dirac fermion systems with tilted cones}

\author{Naoya Tajima$^{1, 2}$}
\author{Takahiro Yamauchi$^1$}%
\author{Tatsuya Yamaguchi$^1$}%
\author{Masayuki Suda$^2$}
\author{Yoshitaka Kawasugi$^3$}
\author{Hiroshi M. Yamamoto$^{2, 4, 5}$}
\author{Reizo Kato$^2$}%
\author{Yutaka Nishio$^1$}%
\author{Koji Kajita$^1$}%

\affiliation{%
$^1$Department of Physics, Toho University - Miyama 2-2-1, Funabashi-shi, Chiba 274-8510, Japan \\
$^2$RIKEN, Hirosawa 2-1, Wako-shi, Saitama 351-0198, Japan \\
$^3$Graduate School of Engineering Science, Osaka University, Toyonaka 560-8531, Japan \\
$^4$Institute for Molecular Science, Okazaki, Aichi 444-8585, Japan \\
$^5$JST-PRESTO, Kawaguchi, Saitama 332-0012, Japan 
}%

\date{\today}

\begin{abstract}
We report the first observation of Shubnikov-de Haas (SdH) oscillations and quantized Hall resistance in the multilayered massless Dirac fermion system $\alpha$-(BEDT-TTF)$_2$I$_3$ with tilted cones. Holes were injected into the thin crystal fixed on a polyethylene naphthalate (PEN) substrate by contact electrification. The detection of SdH oscillations whose phase was modified by Berry's phase $\pi$ strongly suggested that the carrier doping was successful in this system. We succeeded in detecting the quantum Hall effect (QHE) with the steps which is the essence of two dimensional Dirac fermion systems. The number of effectively doped layers was examined to be two in this device. We reveal that the correlation between effective layers plays an important role in QHE.
 \end{abstract}


\maketitle

Since the massless Dirac fermion system was realized in graphene \cite{rf:1, rf:2}, the Dirac particles discovered in other materials, such as graphite \cite{rf:3, rf:4}, bismuth \cite{rf:5}, Fe-based superconductors \cite{rf:6}, topological insulators \cite{rf:7}, and organic conductors \cite{rf:8, rf:9, rf:10, rf:11, rf:12, rf:13, rf:14, rf:15}, have also fascinated physicists. Among them, the organic conductor $\alpha$-(BEDT-TTF)$_2$I$_3$ \cite{rf:16} (Fig. 1(a)) under high pressure offers an ideal testing ground for the Dirac particles. In this paper, we proffer the novel carrier doping technique for the massless Dirac fermion system $\alpha$-(BEDT-TTF)$_2$I$_3$ under high pressure. Using this technique, we first observed Shubnikov-de Haas (SdH) oscillations and quantum Hall effect (QHE) in this system.

$\alpha$-(BEDT-TTF)$_2$I$_3$ is the first bulk (multilayered) two-dimensional (2D) zero-gap conductor with massless Dirac particles. One of the characteristic features of multilayered systems is seen in the inter-layer transport under a transverse magnetic field. In the magnetic field, the energy of Landau levels  in a zero-gap structure is expressed as
\begin{equation}
E_{\rm nLL}=\pm \sqrt{2e \hbar v_{\rm F}^2 |n||B|},  
\label{eqn:1} 
\end{equation}
where $v_{\rm F}$ is the Fermi velocity, $n$ is the Landau index, and $B$ is the magnetic field strength. One important difference between zero-gap conductors and conventional conductors is the appearance of a (n=0) Landau level at zero energy \cite{rf:17}. This special Landau level is called the zero mode. The characteristic features of zero-mode Landau carriers are clearly seen in inter-layer transport \cite{rf:11, rf:18}. In this field configuration, the effect of the magnetic field appears only through the change in the zero-mode density because the interaction between the electrical current and the magnetic field is weak. Thus, remarkable negative (inter-layer) magnetoresistance proportional to the inverse magnetic field was discovered \cite{rf:11, rf:18}. In addition, recent progress in specific heat \cite{rf:19} or NMR \cite{rf:20} studies also revealed characteristic feature of bulk Dirac system. 

\begin{figure}
\includegraphics[viewport = 0 170 700 490, scale=.45, clip]{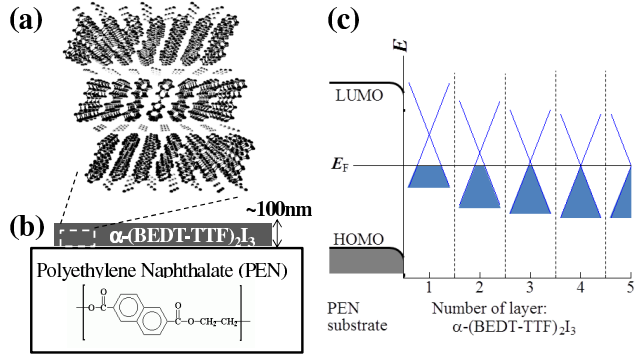}
\caption{\label{fig:1} (color online).  (a) Crystal structure of $\alpha$-(BEDT-TTF)$_2$I$_3$ viewed from the $a$-axis. (b) Schematic diagram of this system. The thickness of the crystal measured with a step profiler was approximately 100 nm. (c) Schematic energy diagram of the present device. The energy spectra (Dirac cones) for each layers are drawn.}
\end{figure}

Another significant feature different from graphene is that the Dirac cones in the present system are tilted, which has been indicated by the tight-binding band calculation \cite{rf:13, rf:14} and the first principles band calculation \cite{rf:15}. Hence, the Fermi velocity, $v_{\rm F}$, is highly anisotropic. The relativistic Landau level structure (Eq. (\ref{eqn:1})) can be corrected, replacing $v_{\rm F}$ with the average $\bar{v}_{\rm F}$. Thus, the tilt of the Dirac cones contributes to the small value of $\bar{v}_{\rm F}$, it was estimated to be approximately $3.5 \times 10^4$ ${\rm ms^{-1}}$ \cite{rf:18}. This value is approximately 1/30 times that of graphene. 

Thus, this material with layered structure and tilted Dirac cones belongs to a broader category of 2D massless Dirac fermion system. However, the relativistic Landau level (Eq. (\ref{eqn:1})) which is the most important characteristic feature in the Dirac fermion system has not directly been obtained yet. Both SdH oscillations and QHE have not been observed yet in this system until now, because the Fermi level is always located at the Dirac point. Moreover, the multilayered structure makes control of the Fermi level by the field-effect-transistor (FET) method much more difficult than the case of graphene. In order to clarify the primitive properties of our Dirac fermion systems, however, the detection of SdH oscillations and QHE are crucial. In this work, we made a breakthrough in the detection of SdH and QHE in the multilayered massless Dirac fermion system $\alpha$-(BEDT-TTF)$_2$I$_3$ under high pressure. 
Our idea is the following.

According to our investigation of the Hall effect in $\alpha$-(BEDT-TTF)$_2$I$_3$, the carrier density at low temperature is only $10^{8}$ cm$^{-2}$/sheet \cite{rf:10}. Yet, the carriers are not localized but mobile with high mobilities. Thus, by fixing a crystal on a substrate weakly negatively charged by contact electrification, the effects of hole doping can be detected in the transport. Indeed, we succeeded in detecting the hole doping effects on the magnetoresistance and the Hall effect by fixing a crystal onto a polyethylene naphthalate (PEN) substrate (Fig. 1(b)). Figure 1(c) is the prospective energy diagram of this device. Holes should be injected into a few layers (pairs of BEDT-TTF molecular layers and I$_3^-$ anion layers; vide infra). From this study, we anticipate that the surface charge produced by contact electrification of the PEN substrate under high pressure is higher than $10^{-3}$ Cm$^{-2}$ at low temperatures.

Single crystals of $\alpha$-(BEDT-TTF)$_2$I$_3$ were synthesized by the electrolysis method \cite{rf:16}. The crystal with a thickness approximately 100 nm was tightly fixed on the PEN substrate by van der Waals forces and/or electrostatic forces, as shown in Fig. 1(b). The resistivities and the Hall resistivities at $p=1.7$ GPa were measured in magnetic fields of up to 7 T at temperatures below 2 K. Experiments were conducted as follows. A sample to which eight electrical leads were attached was put in a Teflon capsule filled with a pressure medium (Idemitsu DN-oil 7373), and then the capsule was set in a clamp-type pressure cell made of the BeCu hard alloy. 


Figure 2(a) shows the magnetic field dependence of the resistance at temperature below 1.7 K.  We notice an oscillation in the resistance at magnetic fields above 2 T. The amplitude of the oscillation increases with decreasing temperature. One vital clue is that the resistance oscillates as a function of $B^{-1}$, as shown in Fig. 2(b).  We succeeded in detecting the SdH oscillations with a frequency of approximately 8.3 T ($=B_0$) in this system for the first time.

\begin{figure}
\includegraphics[viewport = 0 0 700 500, scale=.43, clip]{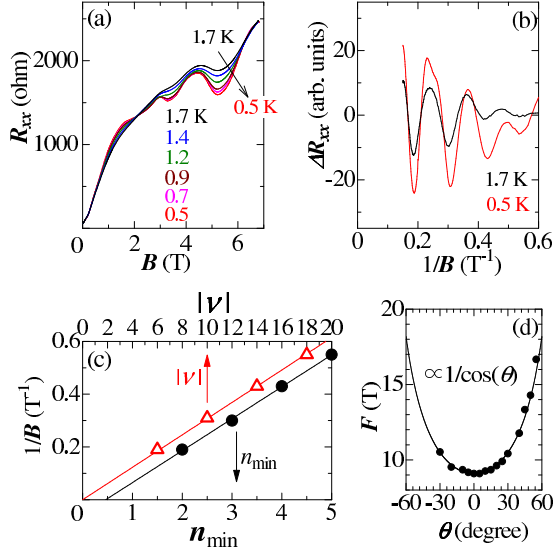}
\caption{\label{fig:2} (color online). (a) Magnetic field dependence of the resistance at temperatures below 1.7 K. Magnetoresistance oscillations are observed. The presence of oscillations with temperature-independent frequencies proportional to $B^{-1}$ shown in (b) strongly suggests that these are SdH oscillations. (b) SdH oscillations as a function of $B^{-1}$ at 0.5 and 1.7 K. (c) The value of $B^{-1}$ for the SdH oscillation minima $n_{\rm min}$ of $\Delta R_{xx}$ (close circle) and for the filling factor $\nu$ (open triangle). (d) Angle dependence of SdH frequencies.}
\end{figure}

We demonstrate that the SdH oscillations in Figs. 2(a) and (b) originate from a 2D Dirac-type energy structure as follows. In the case of a Dirac fermion system, the circular orbit around the Dirac point in the magnetic field would yield Berry's phase $\pi$. This phase would modify the Landau quantization condition so that the zero-mode ($n=0$) Landau level would always appear at the Dirac point. The effect of this phase is further probed in the semi-classical magneto-oscillation description, such as SdH oscillations. The SdH oscillation part of resistivity written by 
\begin{equation}
\Delta R_{xx} =A(B)\cos \left[ 2\pi (B_0/B+1/2+\gamma) \right]
\label{eqn:2} 
\end{equation}
acquires the phase factor $\gamma$=0 or 1/2 for normal electrons with Berry's phase 0 and Dirac particles with Berry's phase $\pi$, respectively \cite{rf:2}. Here, $A(B)$ is the SdH oscillation amplitude. We obtain phase factor $\gamma$ by plotting the values of $B^{-1}$ at the oscillation minima of $\Delta R_{xx}$ (Fig. 2(b)) as a function of their number, as shown in Fig. 2(c). The linear extrapolated value close to 1/2 of the data to $B^{-1}=0$ determines the phase factor, $\gamma$=1/2. Thus, we identify the Dirac particles with a strong 2D nature from the angle dependence of SdH frequencies that obeys the $1/\cos (\theta)$ law, as shown in Fig. 2(d). The polarity of the Hall resistance $R_{xy}$, on the other hand, is positive in the magnetic field region in which SdH oscillations are observed, as shown in Figs. 3(a) and 4(a). Those findings strongly evidence the following. This material fixed on the PEN substrate under high pressure is a 2D massless Dirac fermion system. Though the Dirac cone is tilted, the Landau level structure is expressed as Eq. (\ref{eqn:1}). Moreover, the hole doping in this system is a success. 

Here, we discuss the hole doping level of this device. In the device that uses a multilayered material, carriers are expected to inject into a few layers. According to Horowitz, {\it et al.}, the doped carrier density $n(z)$ in the multilayered materials is written by 
\begin{equation}
n(z) \propto (N_z+L)^{-2},
\label{eqn:3} 
\end{equation}
from the Poisson's equation, where $N_z$ is a number of layers from an interface and $L$ can be regarded as the effective layers (thickness) \cite{rf:21}. The simple SdH signals shown in Fig. 2(a) suggest that \lq\lq the principal effective layer\rq\rq is a single. If several pairs were operated, we would be able to obtain much more complex SdH signals because each doping level would be different (Eq. (\ref{eqn:3})). Performing the second-order differential of $R_{xx}$, however, we find the SdH oscillations with two frequencies as shown in Fig. 3(a). One is the first observed SdH oscillation with $B_0 \sim$ 8.3 T and another one (which is denoted by arrows in Fig. 3(a) ) has a frequency of $B_0 \sim$ 20.4 T. The values of $B^{-1}$ for SdH oscillation minima of $-(d^2R_{xx}/dB^2)$ are shown in Fig. 3(b). The linear extrapolated value of another one close to 1/2 of data to $B^{-1}=0$ strongly indicate that the origin of this SdH oscillation is also Dirac particles. Thus, the holes are injected to more than two layers, and two layers from an interface emitted the SdH oscillations with different frequencies respectively. 

\begin{figure}
\includegraphics[viewport = 0 0 650 470, scale=.43, clip]{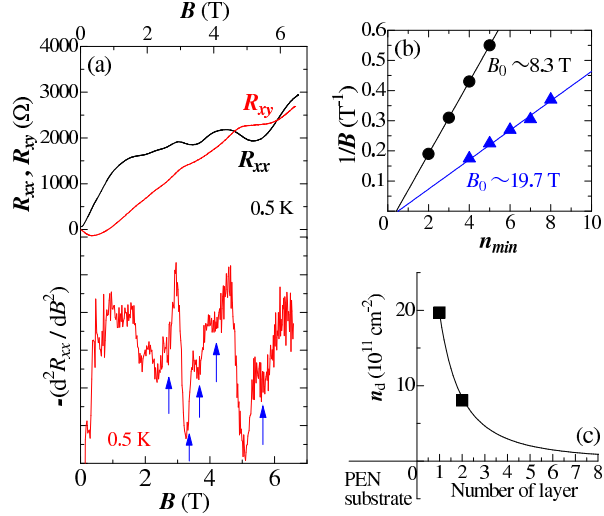}
\caption{\label{fig:3} (color online). (a) Magnetic field dependence of $R_{xx}$, $R_{xy}$and $-(d^2R_{xx}/dB^2)$ at 0.5 K. The arrows shows second Sdh oscillation minima of $-(d^2R_{xx}/dB^2)$. (b) The value of $B^{-1}$ for the SdH oscillation minima $n_{\rm min}$ of $-(d^2R_{xx}/dB^2)$. (c) The distribution of doped carrier density to layered direction. The line is the curve of Eq. (\ref{eqn:3}) with $L=2$. }
\end{figure}

Let us examine the distribution of doped-holes density $n_d$ to layered direction. The carrier density of each layers are estimated to be approximately $19.7 \times 10^{11} {\rm cm^{-2}}$ and $8.3 \times 10^{11}$ ${\rm cm^{-2}}$ from the relationship $n_{\rm d}= B_0 g/\phi_0$, where $g$=4 is the four-fold degeneracy and $\phi_0$=4.14$\times$10$^{-15}$ Tm$^2$ is flux quantum. The Fermi surface areas are approximately 0.05\% and 0.02\% of the first Brillouin zone under ambient pressure. Judging the distribution of the carrier density from Eq. (\ref{eqn:3}), $n_d$ of the first layer from an interface is higher one ($19.7 \times 10^{11}$ ${\rm cm^{-2}}$). Thus, Fig. 3(c) is the distribution of doped carrier density to layered direction in this device. The energy diagram such as Fig. 1(c) is expected. When we assume that the Eq. (\ref{eqn:3}) can be adapted for our system, the best fit of this equation with $L=2$ indicates that the number of effective layer is two. Note that primary (simple) SdH signals shown in Fig. 2(a) are due to the Dirac particles of the second layer because the carrier mobility in the first layer is low. The lower limits of the carrier mobilities in the first and the second layers are roughly examined to be approximately $3.8 \times 10^3$ ${\rm cm^2 V^{-1}s^{-1}}$ and $10^4$ ${\rm cm^2 V^{-1}s^{-1}}$ from the early SdH signals ($\omega_c \tau=(\mu B)^2 >1$), respectively.
Non-smooth surface of PEN will give rise to the low mobility in the first layer.

The detection of the SdH oscillations motivated us to investigate QHE in this system. We show the magnetic field dependence of the Hall resistance $R_{xy}$ at 0.5 K in Fig. 2(a). The two obvious plateaux observed at magnetic field around 3.5 and 5.5 T show that $R_{xx}$ minima are the hallmarks of QHE. We also find the anomalies at fields around 2 and 2.5 T. 

Note that $R_{xx}$ is not zero but shows minima at the plateaux of $R_{xy}$, as shown in Fig. 2. In general complete 2D system cases, $R_{xx}$ should be zero at the quantum Hall plateaux because  $\sigma_{xx}=0$. The present system, however, has a multilayered structure even though few layers, primary the second layer is operated. In this case, the other layers contribute to the conduction so that $R_{xx}$ shows finite values at the quantum Hall plateaux which blurs the SdH oscillations. Moreover, undoped layers also yields the temperature dependence of $R_{xy}$ shown in Fig. 4(a). In this situation, it is difficult to directly demonstrate that the plateaux are complete quantum Hall states. 

The following findings, however, indicate that this is an intrinsic QHE:  (1) The carrier mobility of this material is expected to be higher than $10^5$ ${\rm cm^2 V^{-1}s^{-1}}$ at low temperature. The lower limit of the carrier mobility for the operated layer in this device, on the other hand, is expected to be $\sim 10^4$ ${\rm cm^2 V^{-1}s^{-1}}$ because SdH oscillations are clearly detected around 1 T from the second order differential of $R_{xx}$ as shown in Fig. 3(a). In graphene or oxide heterostructures, QHE has been observed in samples with mobilities lower than those values \cite{rf:1, rf:22}. (2) As mentioned below, the steps of the plateaux are the same as QHE in 2D Dirac fermion systems \cite{rf:1, rf:2}. Especially, the clear plateaux of the Hall conductivity shown in the inset of Fig. 4(a) strongly indicates that those are the hallmark of intrinsic QHE. (3) The breakdown of the plateaux due to the high current intensity is the characteristic feature of QHE \cite{rf:23, rf:24, rf:25}. We verified that the plateaux were blurred with increasing current intensity, as shown in Fig. 4(b). Note that the data of $R_{xy}$ in the normal state are independent of the current intensity. In the quantum Hall state, Joule heat is locally generated in the vicinity of current electrode at the sample edge. Nonequilibrium electron distribution generated at the sample edge gives rise to energy dissipation so that the breakdown of QHE occurs.

\begin{figure}
\includegraphics[viewport = 0 0 700 600, scale=.4, clip]{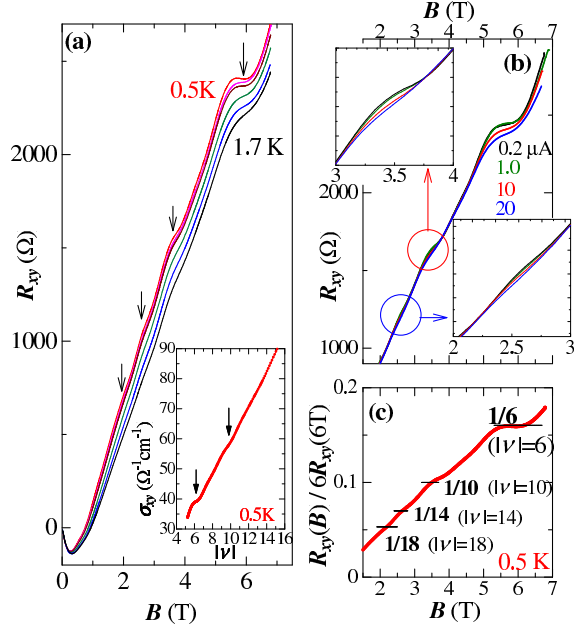}
\caption{\label{fig:4} (color online). (a) Magnetic field dependence of the Hall resistance at temperature below 1.7 K. Arrows indicate the quantum Hall plateaux or anomaly. The inset is $\nu$ dependence of the Hall conductivity $\sigma_{xy}=\rho_{xy}/(\rho_{xx}^2+\rho_{xy}^2)$. The plateaux are clearly observed at $|\nu|=$ 6 and 10. (b) Hall resistance for several current intensities at 0.5 K. The insets show the regions from 2 to 3 T and from 3 to 4 T. The high intensities of the currents blur the plateaux. (c) Magnetic field dependence of the Hall resistance divided by 6 and the value at 6 T. Solid lines show the filling factors for $\nu=$-6, -10, -14, and -18, respectively. }
\end{figure}

Next, we examine the plateaux of $R_{xy}$. In Dirac fermion systems, such as graphene, $R_{xy}$ quantization is in accordance with $R_{xy}^{-1}= \nu \cdot e^2 / h$, where $\nu=\pm g(n+1/2)$ with $g=4$ is the quantized filling factor. A conspicuous effect on the Dirac fermion system is that the factor of half-integer exists. Thus, probes of the quantum Hall plateaux for $|\nu|=$ 2, 6, 10, 14, 18, $\cdots$ are expected. 

The multilayered structure with a few hole-doped layers, however, gives rise to the lack of validity of the examination of $\nu$ from the values of $R_{xy}$ plateau. The values of the $R_{xy}$ plateau depend on the thickness. We cannot ignore the sum of Hall conductivities for the undoped layers. We can see this effect on $R_{xy}$ in magnetic fields below 1.5 T. In this field region, the polarity of $R_{xy}$ is negative and yet holes are injected. This is frequently observed behavior of the present material induced by doped electrons with ppm order that  originate from unstable I$_3^{-}$ anions \cite{rf:6}. 

Thus, we easily examine the filling factors as follows. First, we assume that this material is truly a 2D massless Dirac fermion system, the same as graphene. Second assumption is that the plateaux or anomaly of $R_{xy}$ shown in Figs. 3(a) or 4(a) are the QHE operated on the second layer. The contribution of the first layer is not small. In order to easily examine the filling factor, however, we ignore the Hall conductivity of the first layer here. Based on these assumptions, the filing factors  $\nu=$ -6, -10, -14, -18 are expected at minima of $R_{xx}$ from the SdH oscillations as shown in Fig. 2(c). For example, we can regard the filling factor of the plateau around 5.5 T as $\nu$=-6. When the filling factor of the plateau around 5.5 T is fixed at $\nu$=-6, we find those of the other plateaux or anomaly to be close to $\nu=$-10, -14, and -18, respectively, as shown in Fig. 4(c). These steps are the essence of QHE in 2D Dirac fermion systems. Furthermore, the plateaux of $\sigma_{xy}$ at $\nu=$ 6 and 10 shown in the inset of Fig. 4(a) encourage us that those are the essence. 

Lastly, we briefly mention the bilayer system in our device. In the above discussion, the Hall effect of the first layer was ignored. Holes are injected into the first layer, however, though the mobility is lower than that of the second layer. In the case that the first and the second layers do not depend on each other, the quantum Hall state with a relationship $n_{d1}/n_{d2}=\nu_1/\nu_2$, should be stable at the filling factor of $\nu=\nu_1+\nu_2$, where $n_{d1}$ and $n_{d2}$ are the carrier densities and $\nu_1$ and $\nu_2$ are the filling factors at the first and the second layers, respectively. For example, we obtain $n_{d1}/n_{d2}\sim \nu_1/\nu_2$ with $\nu_1=-14$ and $\nu_2=-6$ at the plateau for $B\sim 5.5$ T. The filling factor of the data at 0.5 K shown in Figs. 3(a) or 4(a), however, lacks validity of this relationship $\nu=\nu_1+\nu_2$, it is $\nu=-11$ ($\neq -14-6$). The consideration of undoped layers strengthens this inconsistency. This discrepancy suggests that the correlation between the first and the second layers plays an important role in the QHE. Recently, Morinari theoretically demonstrated that the correlation between layers narrows and blurs the plateau \cite{rf:26}. In the bilayer device with high carrier mobilities of GaAs, on the other hand, peculiar quantum Hall states associated with the coherence between layers was discovered \cite{rf:27}. Thus, the present device with higher carrier mobilities would provide a new type of quantum Hall state under higher magnetic fields.

In conclusion, we provided crucial evidence that $\alpha$-(BEDT-TTF)$_2$I$_3$ under high hydrostatic pressure is an intrinsic zero-gap conductor with Dirac cone type energy structure. This is the first work that demonstrated the carrier injection effects on the transport phenomena in the bulk zero-gap conductors. We achieved the hole doping into $\alpha$-(BEDT-TTF)$_2$I$_3$ by fixing the thin
crystal to the PEN substrate. The number of effectively doped layers was examined to be two. We were able to detect SdH oscillations with two frequency whose phase are modified by Berry's phase $\pi$. The quantum Hall plateaux for $\nu$=-6, -10, (-14, and -18) due to the primary effective layer was observed. These steps are the essence of QHE in 2D Dirac fermion systems. Moreover, we revealed that the correlation between bilayer plays an important role in QHE.  This device offers an ideal testing ground for the bilayer QHE of the massless Dirac fermion systems.

This work was supported by Grants-in-Aid for Scientific Research (No. 22540379, No.22224006, and No. 23740270) from the Ministry of Education, Culture, Sports, Science and Technology, Japan.


\end{document}